\documentclass[11pt]{article} % use larger type; default would be 10pt

\usepackage[utf8]{inputenc} % set input encoding (not needed with XeLaTeX)

\usepackage{geometry} % to change the page dimensions
\usepackage{graphicx} % support the \includegraphics command and options

\usepackage{booktabs} % for much better looking tables
\usepackage{array} % for better arrays (eg matrices) in maths
\usepackage{paralist} % very flexible & customisable lists (eg. enumerate/itemize, etc.)
\usepackage{verbatim} % adds environment for commenting out blocks of text & for better verbatim
\usepackage{subfig} % make it possible to include more than one captioned figure/table in a single float
\usepackage{authblk}
\usepackage{fancyhdr} % This should be set AFTER setting up the page geometry
\usepackage{sectsty}
\allsectionsfont{\sffamily\mdseries\upshape} % (See the fntguide.pdf for font help)
\usepackage[nottoc,notlof,notlot]{tocbibind} % Put the bibliography in the ToC
\usepackage[titles,subfigure]{tocloft} % Alter the style of the Table of Contents

\newcounter{versionNumber}
\makeatletter 
\AtEndDocument{%
  \immediate\write\@auxout{%
    \string\setcounter{versionNumber}{\number\value{versionNumber}}%
  }%
}
\makeatother
\AtBeginDocument{%
   \stepcounter{versionNumber}
}
\usepackage[calc,useregional]{datetime2}
\begin{document}
%\currenttime
%%% The "real" document content comes below...
\title{Comment on ’Ultradense protium p(0) and deuterium D(0) and their 
relation to ordinary Rydberg matter: a review’ 2019 Physica Scripta 94, 075005
%\\ {\small Version of the document: \theversionNumber} 
%\\ {\small Clock time \currenttime}
}
%\date{} % Activate to display a given date or no date (if empty),
         % otherwise the current date is printed 
\author{Klavs Hansen
\thanks{e-mail: klavshansen@tju.edu.cn, hansen@lzu.edu.cn}}
\affil{ Tianjin University, School of Science, Center for Joint Quantum Studies,China} 
\affil{Lanzhou University, Lanzhou Center for Theoretical Physics, China}
\author{Jos Engelen}
\affil{University of Amsterdam and Nikhef, Science Park 105, 1098 XG Amsterdam, The Netherlands}
\maketitle
\newcommand{\DTMtznow}[2]{%
% store current time in object 'now'
\DTMsavenow{now}%
% convert current time to UTC+0
\DTMtozulu{now}{currzulu}%
% add requested timezone offset to zulu time
\DTMsaveaszulutime{currcest}{\DTMfetchyear{currzulu}}{\DTMfetchmonth{currzulu}}{\DTMfetchday{currzulu}}{\DTMfetchhour{currzulu}}{\DTMfetchminute{currzulu}}{\DTMfetchsecond{currzulu}}{#2}{00}%
% display zulu+offset in requested timezone (= reverse offset)
%\DTMdisplay{\DTMfetchyear{currcest}}{\DTMfetchmonth{currcest}}{\DTMfetchday{currcest}}{}{\DTMfetchhour{currcest}}{\DTMfetchminute{currcest}}{\DTMfetchsecond{currcest}}{#1}{00}%
%}
%test
\DTMdisplay{\DTMfetchyear{currzulu}}{\DTMfetchmonth{currzulu}}{\DTMfetchday{currzulu}}{}{\DTMfetchhour{currzulu}}{\DTMfetchminute{currzulu}}{\DTMfetchsecond{currzulu}}{#1}{00}%
}
% shortcut command for central european summer time (UTC+2)
%\newcommand{\DTMcestnow}{\DTMtznow{+02}{-02}}
%in current timezone: \DTMnow
%in CEST: \DTMcestnow
%GMT (Zulu) \DTMtznow{0}{0}
%{\small \noindent This version generated at clock time \currenttime}\\ \\
{\bf Abstract. }The article by Holmlid and Zeiner-Gundersen (2019 {\it Physica Scripta}
{\bf 94} 075005) contains a number of claims that explicitly or implicitly contradict
fundamental knowledge of modern science. 
Some can only be true if long held conservation laws are broken. 
One such is baryon number conservation. 
A second fatal mistake is the treatment of the structure of molecules that disregard 
fundamental quantum mechanical aspects, such as the concept of kinetic energy 
operators and the Heisenberg indeterminacy relations. 

\section{Introduction}
The article we comment on, which takes the form of a review, contains a large number
of egregious statements and inferences beyond those explicitly addressed here. 
For the sake of space we address in this comment only those that deal with the violation of 
baryon number conservation, and with the treatment of aspects of molecular structure 
that completely ignore the laws of quantum mechanics that govern the structure of matter 
at this scale.
We also include comments on the experimental work used by the authors. 

\section{Baryon number conservation}
Perhaps the most remarkable claim regarding the properties imputed to ultradense 
hydrogen (UDH) is the induction and even spontaneous occurence of nuclear reactions. 
Besides the claim of 'cold fusion' of deuterons in ultra-dense deuterium, protons (p) 
in the UDH are supposed to undergo the reaction
\begin{equation}\sf
pp\rightarrow 3K
\end{equation}
`K' here stands for `kaon', a hadron, which can be either charged or neutral. 
The reaction is energetically possible. 
The mass of a proton is roughly 1 GeV/$\sf c^2$, that of a K roughly 0.5 GeV/$\sf c^2$, 
so there would be a surplus of  0.5 GeV/$\sf c^2$ carried away as kinetic energy that 
is ultimately transferred to the decay products of the K-mesons, notably muons.

But the reaction violates a very fundamental law, namely that of the conservation 
of baryon number or, roughly speaking, that of the conservation of the total number of 
protons and neutrons in the universe. 
A violation of this conservation law has never been observed.
Not in high energy particle physics experiments, nor in any other type of experiment.
It has been the objective of dedicated experiments in large underground facilities 
(shielded from cosmic rays). 
They have all shown no proton decay and have established a lower limit of the 
proton lifetime of 3.6 $\sf 10^{29}$ years. 
That means that no proton was seen to decay in one ton of protons during one year.

The authors suggest that the reaction above can occur as a rearrangement of the quarks 
in the two protons (six in total) into three sets of two quarks (the kaons).
This is completely missing the point that kaons like all other mesons do not consist
of a quark-quark pair but of a quark-antiquark pair. 
In other words, their suggested explanation completely disregards the difference between 
matter and antimatter.
Moreover, kaons contain a type of quark with a property called strangeness and named
strange quarks.
Protons do not, and where the strange quarks in the reaction scheme all of sudden come
from remains unexplained.

Summarizing this part, these claims by the authors are based on words `borrowed' from well 
the well established field of high energy physics, and used completely out of context, 
with no justification and in violation of fundamental laws of nature.

\section{Molecular structure of $\sf H_2$}

As another example of the disregard of well established facts, we want to mention 
the authors' treatment of the structure of the hydrogen molecule.  
This important molecule has been investigated in great detail, both
experimentally and theoretically. 
A number of properties are listed, for example, in the NIST chemistry databook.
The molecule has a ground state bond length of 0.74 \AA. 
Holmlid and Zeiner-Gundersen (in the following HZG) claim the existence of another 
species of the hydrogen molecule with bond length 2.3 pm. 
In their Fig.1 the authors provide their understanding of how this very short bond 
length can come about. 
Their argument is that the molecule has six Coulomb interactions, the two e-e and 
nucleus-nucleus repulsions, and the four e-nucleus interactions (in passing we note 
that the HZG Eq. 1 is incorrect). 
The sum of these six interactions is then claimed to amount to a binding, and that due 
to the increase of the Coulomb interaction with reduced distances, it will render
smaller structures more stable than bigger structures.

This explanation begs the question of why there are any normal $\sf H_2$ molecules.
Somehow there must be a repulsive force at work. 
The answer is very well known and is of course the presence of the kinetic energy 
term for the electrons. 
This is what defines the size of the normal and, we can safely add, to date the only 
observed form of the $\sf H_2$  molecule.

This can be illustrated without going into extensive calculations, even at the level of a 
particle in a box. 
Confining an electron will introduce a kinetic energy that varies with the inverse
square of the size of the box. 
The kinetic energy term will therefore grow faster than the Coulomb terms will 
decrease (go more negative) when the molecular length scale is reduced. 
The two terms strike a balance at one value, and this is at the measured and 
generally accepted 0.74$~\sf \AA$  bond length. 
Although simple, this argument against the simple counting argument of HZG is 
incontrovertible. 
It is based on scaling properties of the potential energy operator, in close parallel 
to the arguments used to derive the virial theorem.
The counting of potential energy contributions of HZG is therefore irrelevant.

The bond length of 0.74 \AA ~can even be understood simpler, as resulting from 
the uncertainty principle, or more specifically as the result of minimizing the total 
energy, which is the sum of the two types of terms:
the kinetic energy of the electrons, and the Coulomb interactions.
(The nuclear kinetic energy can be ignored here). 
The Coulomb interactions are the attractive electron-nucleus interactions, and the 
repulsive electron-electron and nucleus-nucleus interactions. 
Balancing these two then proceeds along the line of the above argument.

The above is an elementary textbook argument, but if more reason is needed 
to understand it, one can consider the hydrogen atom for an even clearer 
demonstration. 
In this atom, there is one potential energy and it is attractive. 
By the argument of HZG, there is therefore nothing to prevent this atom from 
collapsing to a structure where the electron is located on top of the nuclear
charge.
It should not be necessary to explain in any detail why such a suggestion is in
disagreement with all we have learned about the structure of atoms and molecules 
the last 100+ years.
Clearly, the same arguments also raise the question why the 2.3 pm molecules brought 
up by HZG should have any finite bond length at all. 

The treatment of total molecular energies by HZG also includes an argument
derived from considerations of the quantum mechanical wave function. 
The claim is that the counting of potential energy contributions is even more 
favorable than the simple picture in their Fig.1 is suggesting, because the 
electron-electron repulsion supposedly cancels.
The argument of HZG is best given with a quote from their article (p.4, top of
right column): ’With different spin states for the two electrons, they may fill the same
space and one of the repulsive terms (–) disappears effectively.’ What HZG seem to
be saying here is that if the two electrons have opposite spins, the Hamiltonian will be
changed as:
\begin{equation}
\sf
H\rightarrow H-\frac{e^2}{4\pi r_{1,2}}
\end{equation}
That is effectively a postulated interaction of the two electrons based on their spin
projection that will completely cancel the Coulomb interaction between the two. 
It should not be necessary to point out that such interaction has never been observed,
although if present it would be abundantly manifested in atoms, molecules and solid
matter. 
A simple consequence of this suggestion is that the helium atom would have a ground
state energy of exactly 8 times that of the hydrogen atom.
The suggestion seems to be based on a simple misunderstanding of the nature of
a multi-particle wave function (and notably of the Pauli principle). 
The claim is even implicitly contradicted by the authors themselves, who elsewhere 
claim that the postulated H(0) has a spin of 2. 
With a total of four elementary particles in the molecule (two protons and two electrons),
each of spin 1/2, it is impossible for any pairs of those spins, and in particular the spins 
of the electrons, not to be aligned. 
This will render their postulated cancellation of the Coulomb interaction inoperative. 

\section{Experiments, measurements}

Above we have concentrated on some of the extreme claims of the production 
and properties of so called ultra-dense hydrogen. 
These claims are disproportionate, compared to the experimental evidence given. 
The production of UDH is supposed to proceed through a catalytic process employing 
a standard, potasium doped, iron oxide catalyst producing clusters of various sizes of UDH. 
The UDH is then supposed to drip down on a metal surface. 
A common laser \footnote{We quote: `the laser most often used in our experiments
here being a $<$0.5 J Q-switched laser with pulse length in the 5 ns range.'} then 
induces the explosive breakup of these clusters. 
The `reaction products', both charged and neutral, are observed through measurements 
of their time-of-flight. 
The detectors are thin and thick scintillators, some covered with a `catcher foil' (`to detect 
neutrals') placed at two distances from the target. 
A dynode at -7 kV accelerates the positively charged reaction products towards a 
scintillator in front of a photo-multiplier. 
Disentangling the peaks, assigning a mass and a kinetic energy to them is not 
straightforward, very hard to understand and not convincing.
The postulated bond length of a few picometers, in particular, is calculated from the
observation of flight times in mass spectra.
The peaks which, noted in passing, have extremely poor resolution, are assigned without 
any argument to the protons emitted in a Coulomb explosion of the putatively extremely 
strongly bound new form of hydrogen.
No attempt has been documented of any attempt to rule out any other explanation, 
for example the obvious suggestion that the spectra are due to charging up of the 
sample.

Finally, we want to address the stated numbers for the production rates of the 
ionizing radiation. 
The claim on meson production is based on `Time-of-flight current muon signal to two collectors'. 
The analysis of the signal is reported with scant details. 
The interpretation in terms of kaon and pion production and subsequent decay into muons, 
however, is given with great certainty. 
Reactions are even claimed to be also taking place spontaneously, i.e. without exposure of the 
sample to laser light.
Reaction yields are quantified as: `The number of mesons observed in each laser pulse 
is as large as $\sf 10^{15}$, which seems to be the highest meson intensity used anywhere 
in the world.' 
This is a truly astonishing statement for at least two reasons.
One is that this production rate corresponds to an energy output close to 98 kJ with 
an input of 0.5 J laser light. 
Another is that it is made without any reference to radiation protection measures 
that should have been taken.
This type of intensities will cause serious damage to living biological matter in the 
surroundings and even to the experimental equipment used.
 
\section{Final comments}

The paper of Holmlid and Zeiner-Gundersen makes claims that would be truly
revolutionary if they were true.
We have shown that they violate some fundamental and very well established laws in
a rather direct manner. 
We believe we share this scepticism with most of the scientific community. 
The response to the theories of Holmlid is perhaps most clearly reflected
in the reference list of their article. Out of 114 references, 36 are not coauthored by
Holmlid. And of these 36, none address the claims made by him and his co-authors.
This is so much more remarkable because the claims, if correct, would revolutionize
quantum science, add at least two new forms of hydrogen, of which one is supposedly
the ground state of the element, discover an extremely dense form of matter, discover
processes that violate baryon number conservation, in
addition to solving humanity's need for energy practically in perpetuity. 

\subsection*{Acknowledgements}
KH acknowledges support from the National Science Foundation of China with the
grant ’NSFC No. 12047501’ and the Ministry of Science and Technology of People’s
Republic of China with the 111 Project under Grant No. B20063.
\\
Data availability statement: \\
No new data were created or analysed in this study.\\
\subsection*{ ORCID id’s}
Klavs Hansen
https://orcid.org/0000-0001-9746-3711 

\noindent
Jos Engelen
https://orcid.org/0000-0002-9086-8940 See also: $\sf www.nikhef.nl/\sim h02$
\end{document}